\documentclass{PoS}
\usepackage{amsmath}

\title{\vspace{-1.8cm}
Prototype Schwarzschild-Couder Telescope for the Cherenkov Telescope Array: \\ Commissioning
Status of the Optical System}
\ShortTitle{The pSCT for CTA: Commissioning Status of the Optical System}

\author{
C.~Adams~$^{a}$,
G.~Ambrosi~$^{b}$, 
M.~Ambrosio~$^{c}$,
C.~Aramo~$^{c}$,
W.~Benbow~$^{d}$,
B.~Bertucci~$^{be}$,
E.~Bissaldi~$^{fg}$,
M.~Bitossi~$^{h}$,
A.~Boiano~$^{c}$,
C.~Bonavolont\`a~$^{c}$,
R.~Bose~$^{i}$,
A.~Brill~$^{a}$,
J.~H.~Buckley~$^{i}$,
M.~Caprai~$^{b}$,
C.~E.~Covault~$^{j}$,
L.~Di~Venere~$^{g}$,
S.~Fegan~$^{x}$,
\speaker{Q.~Feng}~$^{k}$,
E.~Fiandrini~$^{be}$,
A.~Gent~$^{l}$,
N.~Giglietto~$^{fg}$,
F.~Giordano~$^{fg}$,
R.~Halliday~$^{j}$,
O.~Hervet~$^{m}$,
G.~Hughes~$^{d}$,
T.~B.~Humensky~$^{a}$,
M.~Ionica~$^{b}$,
W.~Jin~$^{n}$,
P.~Kaaret~$^{o}$,
D.~Kieda~$^{p}$,
B.~Kim~$^{q}$,
F.~Licciulli~$^{g}$,
S.~Loporchio~$^{fg}$,
V.~Masone~$^{c}$,
T.~Meures~$^{r}$,
B.~A.~W.~Mode~$^{r}$,
R.~Mukherjee~$^{k}$,
A.~Okumura~$^{s}$,
N.~Otte~$^{l}$,
F.~R.~Pantaleo~$^{fg}$,
R.~Paoletti~$^{ht}$,
A.~Petrashyk~$^{a}$,
J.~Powell~$^{n}$,
K.~Powell~$^{l}$,
D.~Ribeiro~$^{a}$,
J.~Rousselle~$^{w}$,
A.~Rugliancich~$^{h}$,
M.~Santander~$^{n}$,
R.~Shang~$^{q}$,
B.~Stevenson~$^{q}$,
L.~Stiaccini~$^{ht}$,
L.~P.~Taylor~$^{r}$,
L.~Tosti~$^{be}$,
V.~Vagelli~$^{bev}$,
M.~Valentino~$^{uc}$,
J.~Vandenbroucke~$^{r}$,
V.~Vassiliev~$^{q}$,
P.~Wilcox~$^{o}$,
D.~A.~Williams~$^{m}$, 
for the CTA SCT Project
\thanks{for consortium list see PoS(ICRC2019)1177}
\\
\llap{$^a$} Physics Department, Columbia University, New York, NY 10027, USA\\
\llap{$^b$} INFN Sezione di Perugia, Perugia, Italy\\
\llap{$^c$} INFN Napoli, Italy\\
\llap{$^d$} Center for Astrophysics | Harvard \& Smithsonian, Cambridge, MA 02138, USA\\
\llap{$^e$} Universit\`a degli Studi di Perugia, Perugia, Italy\\
\llap{$^f$} Dipartimento Interateneo di Fisica dell'Universit\`a e del Politecnico di Bari\\
\llap{$^g$} INFN Bari, Via E. Orabona 4, 70125 Bari, Italy\\
\llap{$^h$} INFN Sezione di Pisa, Pisa, Italy\\
\llap{$^i$} Department of Physics, Washington University, St. Louis, MO 63130, USA\\
\llap{$^j$} Department of Physics, Case Western Reserve University, Cleveland, Ohio 44106\\
\llap{$^k$} Department of Physics and Astronomy, Barnard College, Columbia University, NY 10027, USA\\
\llap{$^l$} School of Physics \& Center for Relativistic Astrophysics, Georgia Institute of Technology, 837 State Street NW, Atlanta, GA 30332-0430, USA\\
\llap{$^m$} Santa Cruz Institute for Particle Physics and Department of Physics, University of California, Santa Cruz, CA 95064, USA\\
\llap{$^n$} Department of Physics and Astronomy, University of Alabama, Tuscaloosa, AL 35487, USA\\
\llap{$^o$} Department of Physics and Astronomy, University of Iowa, Iowa City, IA 52242, USA\\
\llap{$^p$} Department of Physics and Astronomy, University of Utah, Salt Lake City, UT 84112, USA\\
\llap{$^q$} Department of Physics and Astronomy, University of California, Los Angeles, CA 90095, USA\\
\llap{$^r$} Department of Physics and Wisconsin IceCube Particle Astrophysics Center, University of Wisconsin, Madison, WI 53706, USA\\
\llap{$^s$} Institute for Space--Earth Environmental Research and Kobayashi--Maskawa Institute for the Origin of Particles and the Universe, Nagoya University, Nagoya 464-8601, Japan\\
\llap{$^t$} Dipartimento di Scienze Fisiche, della Terra e dell'Ambiente, Universit\`a degli Studi di Siena, Siena, Italy\\
\llap{$^u$} CNR-ISASI, Italy\\
\llap{$^v$} Now at ASI Italian Space Agency - Scientific Research Unit, Roma, 00133, Italy\\
\llap{$^w$} Subaru Telescope NAOJ, Hilo HI 96720, USA\\
\llap{$^x$} LLR/Ecole Polytechnique, Route de Saclay, 91128 Palaiseau Cedex, France \\
\newline
E-mail: \email{qifeng@nevis.columbia.edu}}

\abstract{
The Cherenkov Telescope Array (CTA), with more than 100 telescopes, will be the largest ever
ground-based gamma-ray observatory and is expected to greatly improve on both gamma-ray
detection sensitivity and energy coverage compared to current-generation detectors. The 9.7-m
Schwarzschild-Couder telescope (SCT) is one of the two candidates for the medium size
telescope (MST) design for CTA. The novel aplanatic dual-mirror SCT design offers a wide
field-of-view with a compact plate scale, allowing for a large number of camera pixels that improves
the angular resolution and reduce the night sky background noise per pixel compared to the
traditional single-mirror Davies-Cotton (DC) design of ground-based gamma-ray telescopes. The
production, installation, and the alignment of the segmented aspherical mirrors are the main
challenges for the realization of the SCT optical system. In this contribution, we report on the
commissioning status, the alignment procedures, and initial alignment results during the initial commissioning phase of the optical system of the prototype SCT. 
}

\FullConference{36th International Cosmic Ray Conference -ICRC2019-\\
		July 24th - August 1st, 2019\\
		Madison, WI, U.S.A.}

\begin{document}
\addtocounter{page}{2}
\section{A brief introduction of the optical alignment systems of the pSCT}

The design of the optical system of the prototype Schwarzschild-Couder Telescope (pSCT) is derived from
the exact Schwarzschild aplanatic solution described in \cite{Vassiliev07}. 
The small pixel scale and superior off-axis optical point spread function (PSF) are two of the main advantages of the SCT design, allowing for an improvement of the imaging capability of air showers over the one-mirror Davies-Cotton design \cite{DaviesCotton57} by about an order of magnitude. Such an improvement enables important investigations including detailed morphology studies of extended TeV gamma-ray sources, which provide insights into the acceleration and transport of Galactic cosmic rays. To achieve the designed optical PSF of the pSCT, the alignment of the optical system is the key issue. 

The 9.7-m diameter primary mirror of the pSCT consists of two rings, an inner ring with 16 pentagonal mirror panels, and an outer ring with 32 quadrilateral panels. The 5.4-m diameter secondary mirror consists of an 8-panel inner ring, and a 16-panel outer ring, with geometry similar to that of the primary mirror. 
All the mirror panels were installed by 2018 December. 

The mirror surfaces of the pSCT were briefly uncovered during the inauguration and the first light of the pSCT in 2019 January. However, they have remained covered until 2018 July, to allow safe construction activity of the auxiliary system. 
Hardware challenges after installation have been overcome, and the optical system of the pSCT is currently fully functional.  
In the meantime, the commissioning activity for the optical alignment system has started, employing strategies without using the optical surfaces. 

In order to achieve an ideal optical PSF ($\sim$3.5'--4'), the alignment precision of the pSCT is required to be as good as 0.1~mrad for the relative tip and tilt between panels and 1.1~mm for the translation (perpendicular to the optical axis) of panels \cite{Nieto15,Rousselle13}. The most strict requirement comes from the tip-tilt of the primary panels, allowing a maximum panel-to-panel misalignment of $\sim$100~$\mu$m when it is entirely due to tip-tilt misalignment.

The optical alignment system consists of two subsystems. A panel-to-panel alignment system aligns every panel with its neighbors. A global alignment system aligns the entire primary and secondary mirror, as well as the gamma-ray camera, with respect to each other, and monitors their relative positions. 
Initial calibration of the alignment system components has been performed in the lab before their installation.

Both the deformation of the optical support structure (OSS) due to changing gravitational loads at different elevation angles and the thermal expansion of the OSS due to seasonal temperature changes can lead to non-negligible motion of the panels. 
After initial alignment, 
the response of the optical alignment system to these effects will be calibrated by measuring and recording the relative positions between optical components at different pointing directions and temperatures, allowing offline pointing corrections. The optical components will be realigned if the measured misalignment is large enough to affect the size of the optical PSF.  
The initial alignment strategies and the testing results of the pSCT, without using its optical surfaces at the current starting phase of the commissioning will be briefly described below. 
In the future, optical methods using images of stars and additional calibration light sources will be implemented, to allow a measurement of the optical PSF and alignment strategies directly related to it. The optics team of the pSCT is actively developing alignment software, including a prototype
engineering graphical user interface (GUI), towards the full commissioning alignment of the pSCT. 


\begin{figure}
    \centering
  \begin{minipage}[b]{0.48\textwidth}
    \includegraphics[width=\textwidth]{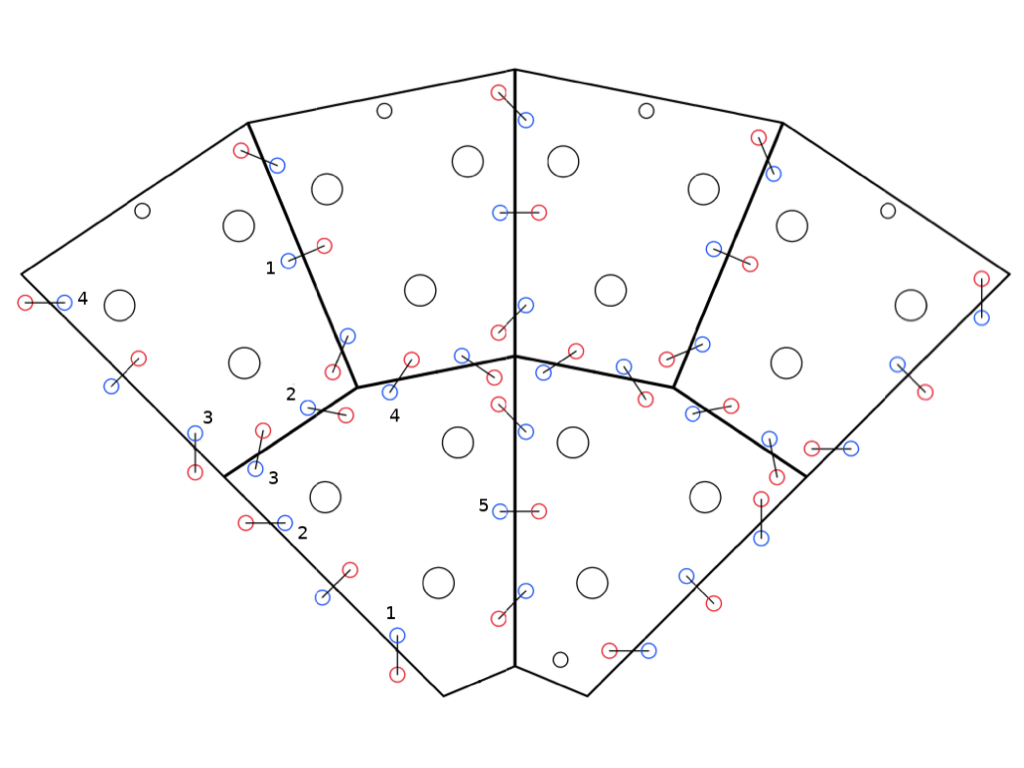}
\end{minipage}
\begin{minipage}[b]{0.48\textwidth}
     \strut\vspace*{-\baselineskip} \includegraphics[width=\textwidth]{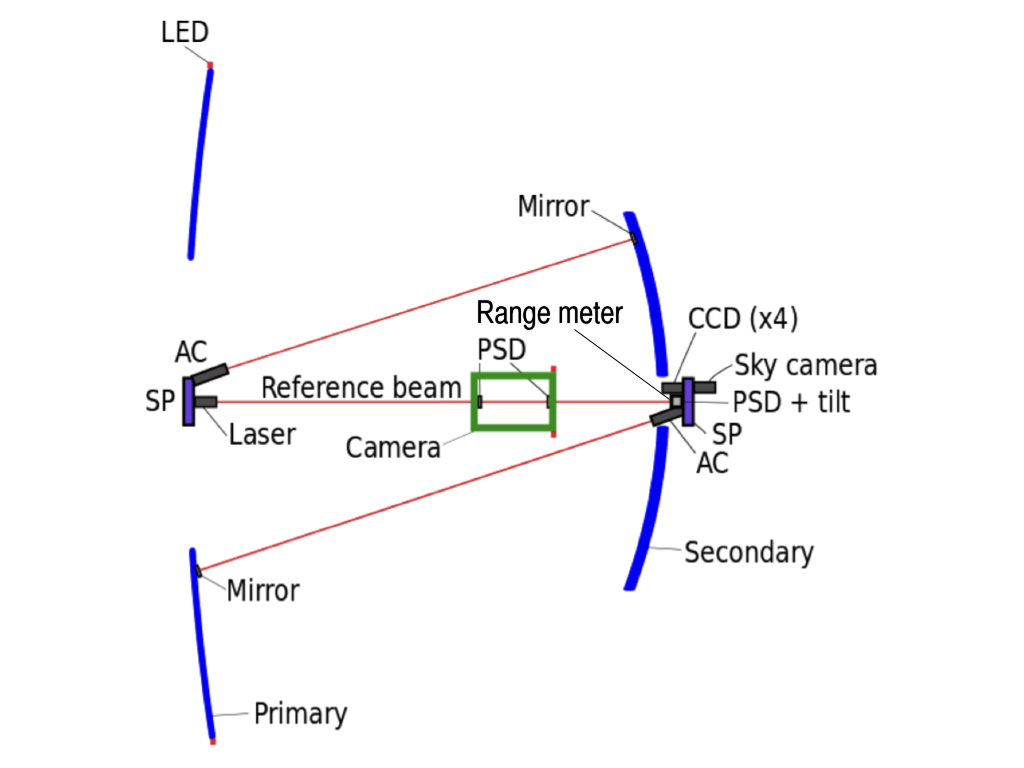}
\end{minipage}
    \caption{{\it Left:} Illustration of two sectors of the secondary mirror panel modules of the pSCT. Blue and red circles indicate the position of the MPESs. 
    {\it Right:} Illustration of global alignment system of the pSCT, taken from \cite{Nieto15}.}
\end{figure}
\label{fig:sector}


\section{Panel-to-panel alignment system of the pSCT} 

Each mirror panel of the pSCT is supported by six actuators, the length of each of which can be changed in 3-$\mu$m increments, forming a Stewart Platform \cite{Sreenivasan94}. By extending or retracting the six actuators of a mirror panel through software, its six degrees of freedom (DoF; three for center of mass and three for normal direction) can be exactly controlled. 
For an edge between two panels, two (for edges between an inner and an outer panel) or three (for edges between two inner or two outer panels) mirror panel edge sensors (MPESs) are installed (see~Figure~\ref{fig:sector} for an illustration of the MPES layout). An MPES consists of a laser and a CCD camera, each mounted near the edge on one of the neighboring panels. The centroid (two DoF) of the laser image on the CCD camera is measured with a $\sim$10~$\mu$m resolution and a $\sim1$~cm range, enabling us to measure the relative alignment of two panels with a comparable precision and range. 

With the optical surfaces covered, the panel-to-panel alignment during the early commissioning of the pSCT relies solely on the measurements from MPESs. 
The motion of all 72 panels on the two mirrors of the pSCT consists of 432 DoF. A total of 312 MPESs, which provides 624 measurement values for the edges between panels (greatly exceeding the total number of DoF of panel motion), allow full constraints of the panel-to-panel alignment. The extra constraints offered by MPESs allow for the margin of unexpected hardware failure. 

Linear algebra can be employed to describe the changes in MPES measurements along an edge from any small motion of a relevant panel, and to solve the motions of panels that bring edges into alignment. 

Let $j$ and $k$ denote two neighboring panels with arbitrary lengths of actuators $\vec{L}_j$ and $\vec{L}_k$, respectively. The alignment of the edge between these two panels is measured as the laser centroid coordinates $\vec{\sigma}_{j,k}$ in the MPESs. 
We define the response matrix $R_{j,k} = \delta \vec{\sigma}_{j,k} / \delta \vec{L}_{k}$, so that it transforms small displacements of the actuators $\delta \vec{L}_{k}$ of a panel $k$ to the corresponding changes of the MPES readings for the edge between panels $j$ and $k$, i.e., $\delta \vec{\sigma}_{j,k} = R_{j,k} \delta \vec{L}_{k}$. 

Once the reference MPES coordinates $\vec{\sigma}_{j,k}^\text{ref}$ when the two panels are perfectly aligned with respect to each other are known, then one can express the relative misalignment of an edge as the difference between the reference and actual MPES coordinates $\Delta \vec{\sigma}_{j,k} = \vec{\sigma}_{j,k}^\text{ref} - \vec{\sigma}_{j,k}$. The task of panel-to-panel alignment then becomes finding a set of $\delta \vec{L}_{k}$, which defines the motions for all panels $\forall k$ that lead to ideal alignment $\Delta \vec{\sigma}_{j,k}=0$ for all edges $\forall j,k$. 

Before the panels were installed onto the OSS of the pSCT, every edge between two adjacent panels was aligned on an optical table in the lab with the help of a coordinate measuring machine. Then the MPESs for each aligned edge were installed, and the reference MPES coordinates $\vec{\sigma}_{j,k}^\text{ref}$ were recorded. Finally, the response matrix $R_{j,k}$ (and $R_{k,j}$) was measured by extending and retracting each actuator for panel $k$ (and $j$) and recording the change in MPES readings $\delta \vec{\sigma}_{j,k}$. 



To formulate panel-to-panel alignment, we start with the example of moving only one inner (or outer) panel $k$ to align to a neighboring inner (or outer) panel $j$. Such an edge is equipped with three MPESs, therefore $\delta \vec{\sigma}_{j,k}$ provides 6 constraints. 
Solving the equation 
$R_{j,k} \delta \vec{L}_{k} = \Delta \vec{\sigma}_{j,k}$  
and moving panel $k$ by $\delta \vec{L}_{k}$ would align the edge $j,k$. 

Another example is to align a three-panel sector including an inner panel $m$ and its two neighboring outer panels $j$ and $k$. 
The procedure involves moving any two of these panels (12 DoF), e.g., $j$ and $k$, to align the seven MPESs (14 constraints), formulated as follows: 
\begin{equation}
[(R_{j,k} \delta \vec{L}_{k} + R_{k,j} \delta \vec{L}_{j}), R_{m,j} \delta \vec{L}_{j},  R_{m,k} \delta \vec{L}_{k} ] = [\Delta \vec{\sigma}_{j,k}, \Delta \vec{\sigma}_{m,j}, \Delta \vec{\sigma}_{m,k}], 
\label{eq3}
\end{equation}
where brackets 
denote the concatenation of vectors. Note that the 12-DoF equation is overdetermined by the 14 MPES readings, and therefore the solutions of the alignment panel motions can always been found. 
The above sector alignment can be expanded to the alignment of a sector of an arbitrary number of panels, and can also take into consideration the measurement uncertainties of the reference MPES readings and response matrices. 

Software development for the panel-to-panel alignment following the above methodology is now in progress for the past few years, and many alignment methods have been implemented. Software development will continue to be one of the main commissioning tasks of the pSCT. The large number of devices presents challenges for the alignment software. Currently, software is being developed to parallelize all measurements and calculations, to speed up the alignment process, upgrade the client to overcome the operating system limit on the number of client-server connections, and prototype an engineering GUI to make alignment more user friendly. 

The web-based prototype GUI software will make it easier for users to monitor the status of the optical system and control the optical alignment. The GUI is responsible for periodic updating of information for the status of the optical components (e.g. temperatures, MPES readings and actuator lengths). The GUI should also provide clear and sufficient real-time feedback including warnings for illegal actions, alerts for critical errors, and software limits on panel motions to prevent panel collision. Finally, the GUI should allow multiple levels of access (observer, engineer, and super-user) to restrict user actions to ensure the safety of the optical system. 


\section{Global alignment system of the pSCT} 

The right panel in Figure~\ref{fig:sector} illustrates the global alignment system (GAS) of the pSCT as introduced in \cite{Nieto15}. 
Two optical tables (OTs) at the centers of the primary (OT1) and secondary (OT2) mirrors, respectively, are equipped with optical devices to achieve global alignment between the primary and secondary mirrors, as well as the gamma-ray camera. 
Each OT is supported by a Stewart Platform (SP) identical to those used for mirror panels, allowing control of their full motions through software. The gamma-ray camera is mounted on a movable inner structure enclosed in the housing structure fixed to the OSS. The camera motion along the optical axis and its tip/tilt are actively controlled by three stepper motors (one on top and two on the bottom) parallel to the optical axis. The other three DoF of the camera can only be adjusted manually. 


A reference laser beam from OT1 is measured by position sensitive devices (PSDs) in the optical module at the center of the gamma-ray camera and OT2, and these three optical components can be aligned with respect to the reference laser beam, defining the optical axis. The distances between OT1, OT2, and the gamma-ray camera are measured by a range meter on OT2. 
As the telescope changes its pointing direction, the bending and sagging of the structure may change the relative position between these three optical components. OT1 and OT2 will then be moved into alignment with the camera, maintaining the optical axis. 

The relative shift between the optical axis and the mirror panels will be monitored, providing information for pointing correction. 
Three CCD cameras on each OT image three panels of the opposite mirror (referred to as GAS panels), and measure their relative position with respect to the OT through image analysis of the six LEDs (the positions of which are measured with a 10~$\mu$m precision in the lab) mounted on each GAS panel. These measurements offer sufficient constraints on the translation of the panels, with the highest precision for the two DoF perpendicular to the optical axis. 
To improve the measurement of tip-tilt (two DoF) of the GAS panels with respect to the OTs to $\sim$20~$\mu$rad accuracy, an autocollimator (AC) on each OT and a retroreflector on a corresponnding GAS panel are installed. 


A sky-facing CCD camera on the back of OT2 images a $4.4^\circ \times 3.3^\circ$ star field at a given pointing, using which the center of the field-of-view (FoV) can be calculated from astrometry. 
Pointing calibration can then be performed by pointing the telescope at a star, and comparing the images of the sky taken by the sky CCD camera with the images of the focal plane taken by a GAS gamma-ray CCD camera. The pointing given by the telescope tracking system can also be used for calibration. 
The calibration results will be used to develop a telescope bending model. 
Pointing correction can be done with all observations using the calibration information and images from the sky CCD camera during operation.


\section{Alignment status of the pSCT}

\begin{figure}[h]
\vspace*{-\baselineskip}
\centering
    \includegraphics[width=0.43\textwidth]{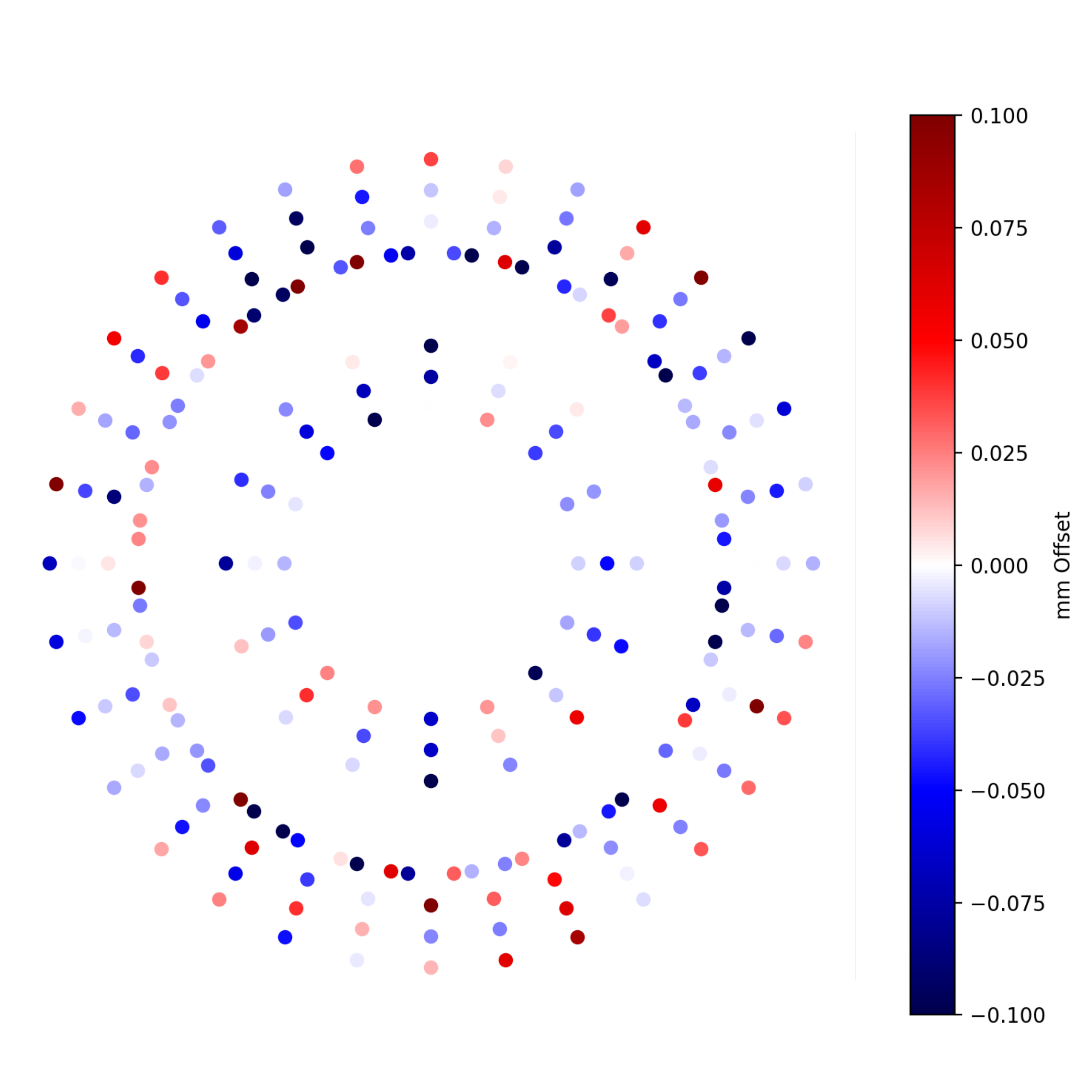}
    \includegraphics[width=0.51\textwidth]{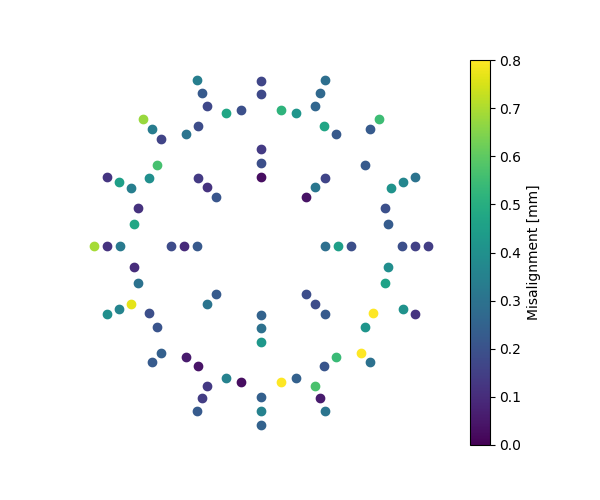}
    \caption{
    Testing of the changes in panel-to-panel alignment of the pSCT under different conditions without moving the optical components. The differences in the panel-to-panel alignment of the primary mirrors ({\it left}) were calculated from two readings of the MPES measurements nine hours apart at 10$^\circ$C and 20$^\circ$C, when the pSCT was parked at $20^\circ$ elevation. The ongoing commissioning alignment for the secondary mirrors, with a mean and standard deviation of the misalignment being $0.3\pm0.2$~mm ({\it right}). 
    }
\label{fig:align}
\end{figure}

Since the successful inauguration of the pSCT in 2019 January, one of the main commissioning goals of the pSCT project is to achieve both panel-to-panel and global alignment of the optical system. 
The initial panel-to-panel alignment 
process is in progress. The edge misalignment was smaller than 0.8~mm for all of the secondary MPESs as of 2019 July (see~Figure~\ref{fig:align} right panel). 
Software development with more efficient parallel measurement and calculation, as well as alignment iterations using the prototype alignment software are being carried out, and significant improvement of the panel-to-panel alignment of the pSCT towards the ideal $\sim$100~$\mu$m precision is expected in the near future. 


To estimate the expected range of motions from telescope bending, the deformation of the OSS under gravity at different elevations before the installation of mirrors are measured, and the bending of the OSS at a few primary outer panel locations was found to be larger than the values from the OSS design simulations ($\sim$1~mm). The initial bias alignment (both panel-to-panel and global) of the optical system is being carried out at an elevation angle of 60$^\circ$. 
After the initial panel-to-panel alignment, we have been testing the stability of the alignment at different temperatures (about 10$^\circ$C to 25 $^\circ$C), as well as at different elevation angles (20$^\circ$ and 60$^\circ$) without moving any panels. The drift in the MPES readings due to temperature variation are found to be under 0.3~mm, as shown in Figure~\ref{fig:align}. 

\begin{figure}[t]
\vspace*{-\baselineskip}
    \centering
      \begin{minipage}[b]{0.48\textwidth}
    \includegraphics[width=\textwidth]{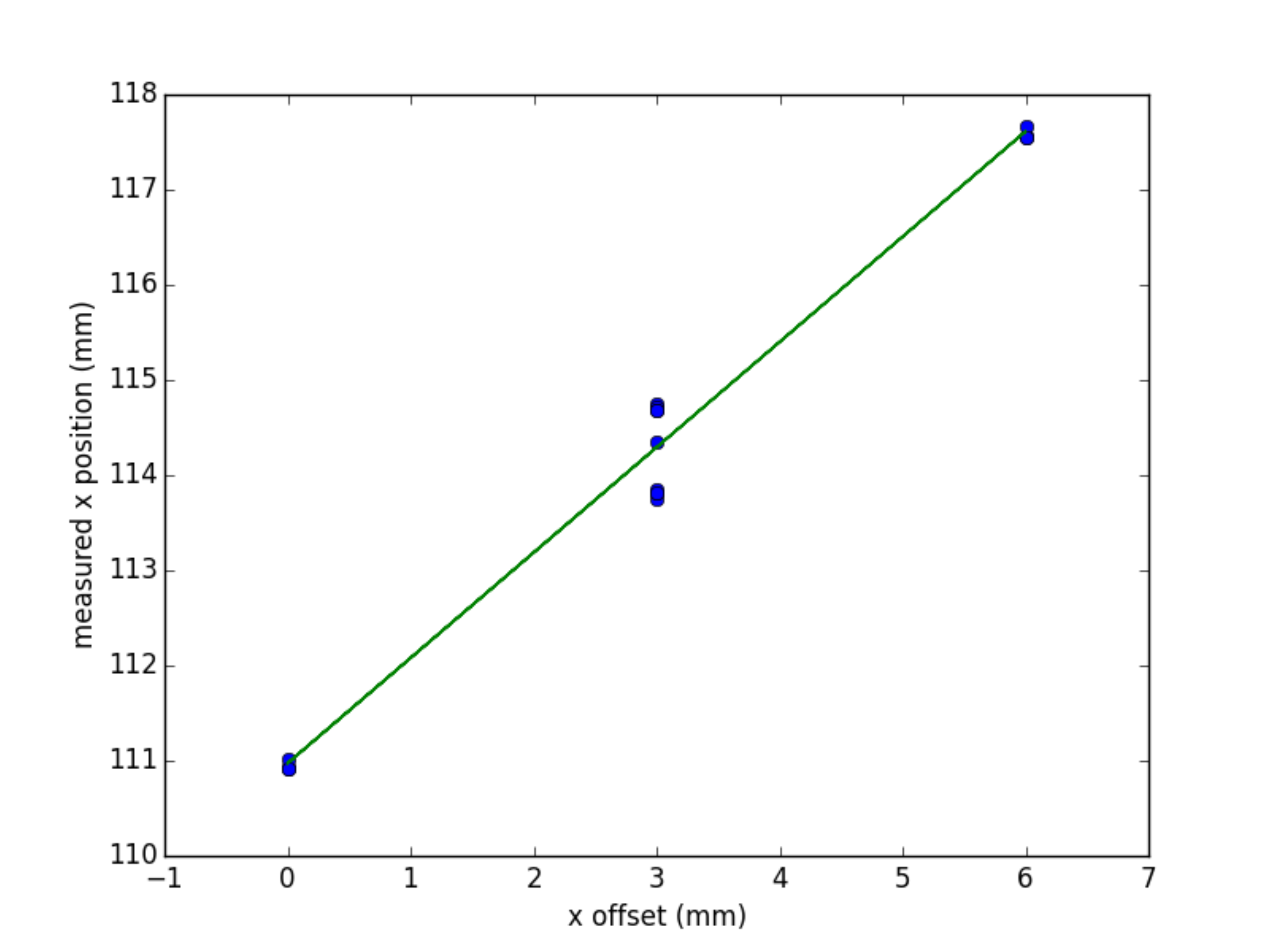}
\end{minipage}
\begin{minipage}[b]{0.39\textwidth}
     \strut\vspace*{-1.\baselineskip} \includegraphics[width=\textwidth]{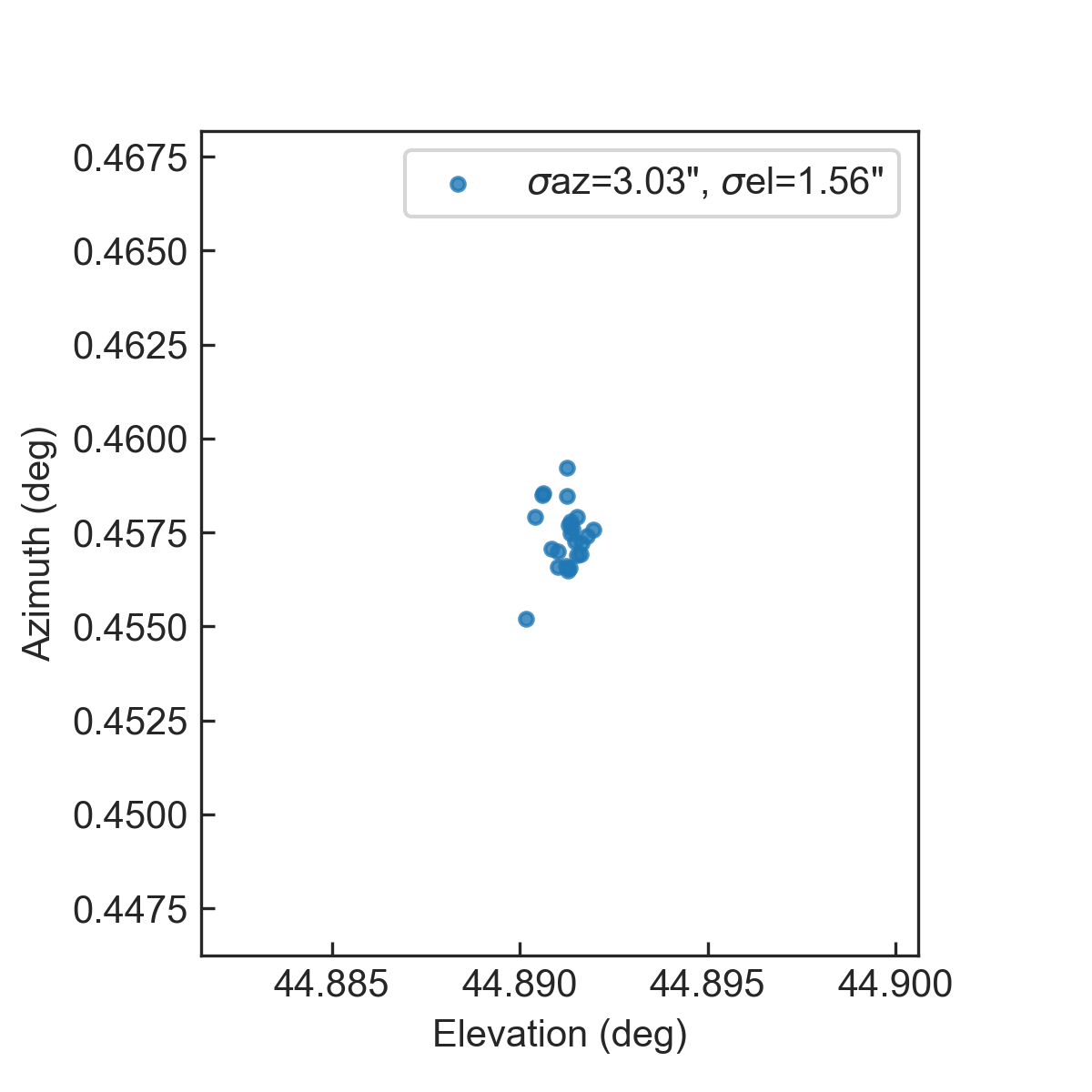}
\end{minipage}
    \caption{{\it Left:} Testing results to recover panel translation along one direction ($x$) using 18 images taken from a GAS CCD camera. 
    {\it Right:} Testing the stability of the astrometry solution of the center of FoV of the sky CCD camera, using 23 images when the telescope was parked at roughly an azimuth angle of 0$^\circ$ and an elevation angle of 45$^\circ$.}
\label{fig:GAS}
\end{figure}

The GAS system has been fully installed, and is currently going through a calibration and commissioning phase. 
Test results were obtained from a GAS CCD camera and the sky camera using prototype software, as shown in Figure~\ref{fig:GAS}. 
To test the ability of measuring the relative translation of a GAS panel with respect to an OT, we moved a test GAS panel along a given axis by $\pm$3~mm repeatedly, and took 18 images using the corresponding GAS CCD camera. The reconstructed panel translations were accurate within 0.25~mm (see~Figure~\ref{fig:GAS} left panel). 
To test the reproducibility of the calculation of the center of FoV of the sky CCD camera, we took 23 images at 0.5~s exposure of drifting star fields over a few hours when the telescope was parked at 45$^\circ$ elevation. The standard deviation of the reconstructed sky CCD camera pointing is below 3$"$ (see~Figure~\ref{fig:GAS} right panel). 

The camera of the pSCT was initially aligned to the center of the mechanical structure in 2019 July, and moved into the focal plane using the range meter on OT2. 
The main focus in the commissioning phase of the GAS system will be software development utilizing all GAS components, as well as the images of bright stars and external calibration light sources after the optical surfaces are uncovered. The effect of telescope bending on the PSF size and offset, as well as any potential hysteresis associated with it, will be investigated. 

\section{Summary and outlook}
The completion of all construction activities of the pSCT has been achieved in 2018 December, followed by its successful inauguration in 2019 January. 
As of 2019 June, all components of the optical system are functional.
Good progress is being made on the initial alignment with optical surface covered.  
Entering the phase of optical alignment, future work will be done to characterize the deformations of the OSS, as well as the on-axis and off-axis optical PSF.
Software implementation of all alignment methods will continue to be carried out, maintaining excellent PSF and measuring pointing corrections in real time during observations of the pSCT. 


\bibliographystyle{elsarticle-num}
\bibliography{ref.bib}

\end{document}